\documentclass[useAMS, usenatbib, usegraphicx]{mn2e}
\usepackage{amssymb}
\usepackage{booktabs}
\newcommand\ApJ{ApJ}

\newcommand\AnA{A\&A}

\def\alf{Alfv\'en\,}
\def\alfc{Alfv\'enic\,}
\def\bq{\begin{equation}}
\def\eq{\end{equation}}
\def\ee #1 {\times 10^{#1}}
\def\ut #1 #2 { \, \rmn{#1}^{#2}}
\def\u #1 { \, \rmn{#1}}

\let\grad=\nabla
\newcommand\cross{\bmath{\times}}
\newcommand\etaH{\eta_\mathrm{H}}

\def\curl{{\grad \cross}}
\def\div #1 {\grad \cdot #1}

\def\alpi1{\alpha_{1i}}
\def\alpi2{\alpha_{2i}}
\def\hB{\hat{\bmath b}}
\def\vB{\bmath{v}_B}
\def\Jpa{\bmath{J_\parallel}}
\def\v{\bmath{v}}
\def\dv{\delta \v}

\def\U{\bmath{U}}
\def\v{\bmath{v}}
\def\vi{\bmath{v}_i}

\def\ve{\bmath{v}_e}
\def\vi{\bmath{v}_i}

\def\vn{\bmath{v}_n}
\def\va{v_A}
\def\J{\bmath{J}}
\def\B{\bmath{B}}
\def\dB{\delta \B}

\def\E{\bmath{E}}
\def\dE{\delta \E}
\def\dEx{\delta E_x}
\def\dEy{\delta E_y}
\def\J{\bmath{J}}

\def\dv{\bmath{\delta\v}}
\def\dvx{\delta v_x}
\def\dvy{\delta v_y}
\def\dE{\bmath{\delta\E}}
\def\dB{\bmath{\delta\B}}

\def\rhon{\rho_{in}}
\def\rhox{\rho_{ex}}

\newcommand{\delt} [1] {\frac{\partial #1}{\partial t}}

\def\oma{{\omega_A}}
\def\b1{{\bar{\omega}}}
\def\bo2{\bar{\omega}^2}
\def\L1{{\bf {\it L}}}

\title{Surface waves in the partially ionized solar plasma slab}
\author[B.P.Pandey]
        {B.P. Pandey
\thanks{E-mail:birendra.pandey@mq.edu.au}\\
\\
{Department of Physics, Astronomy \& Research Centre for Astronomy, Astrophysics \& Astrophotonics,}\\Macquarie University, Sydney, NSW 2109, Australia}
\date{\today}
\pagerange{\pageref{firstpage}--\pageref{lastpage}}
\pubyear{2013}
\begin{document}
\maketitle
\label{firstpage} 
\begin{abstract}
\maketitle
The properties of surface waves in the partially ionized, incompressible magnetized plasma slab are investigated in the present work. The waves are affected by the non—-ideal MHD effects which causes the finite drift of the magnetic field in the medium. When the finite drift of the magnetic field is ignored, the characteristics of the wave propagation in the partially ionized plasma fluid is similar to the ideal MHD except now the propagation properties depend on the fractional ionization of the medium. 

In the presence of Hall diffusion, the propagation of the sausage and kink surface waves depends on the level of fractional ionization of the medium. For example, short wavelength surface modes can not propagate in the medium if the scale over which Hall operates is comparable to the size of the plasma slab. With the increasing ionization, the surface modes of shorter wavelength are permitted in the system.    

When both the Hall and Pedersen diffusion are present in the medium, the waves undergoes damping. In case of Pedersen dominating Hall, the damping of the long wavelength fluctuations are dependent on the ratio of the plasma densities inside and outside the slab and on the square of the Pedersen diffusivity. For typical solar parameters, waves may damp over few minutes. 
\end{abstract}

\section{Introduction}
The ideal magnetohydrodynamics (MHD) provides the most basic framework in which the dynamics of the magnetized and highly stratified solar plasma is generally investigated \citep{P87, GP04, A09}. In this framework the magnetic, inertial and pressure forces interact with each other within the perfectly conducting plasma environment. However, the ideal MHD has its own range of validity set mainly by the relevant length and time scales of the system under study \citep{F82}. The typical length scale of the system under study must be much larger than the Larmor radius of the plasma particles, while the temporal scale needs to be much shorter than the time scale of the collisions among the plasma particles. Therefore, the context of the ideal MHD is very restrictive to the solar atmosphere considering that the photosphere-—chromosphere is weakly ionized with the overwhelming presence of the neutral hydrogen. In addition, reconnection of the magnetic field lines e.g., in the flares can only proceed in the presence of non--ideal MHD effects. The ideal MHD breaks down not only in the lower atmosphere [$\lesssim 2.5\,\mbox{Mm}$, \cite{P08b, PW13}] but also in the coronal patches where filaments of weakly ionized matter is observed \citep{S09}. Clearly, the extrapolation of the ideal MHD framework to the relatively cold chromosphere—-photosphere and its extension to the footpoint motion of the magnetic flux tubes does not represent the ambient physical reality. Furthermore, the collision dominated solar atmosphere may be unstable  to the fast growing non-—ideal MHD instabilities in the presence of convective shear flows \citep{PW12, PW13}. Therefore, the description of the lower solar atmosphere requires a paradigm shift from the fully ionized, ideal MHD framework to the partially ionized non—-ideal MHD framework to investigate the low frequency behaviour of the medium.  

However, notwithstanding the presence of partially ionized lower solar atmosphere, while dealing with the long wavelength ($\sim$ several pressure scale height) fluctuations, the photosphere—-chromosphere region can be treated as an infinitesimally thin boundary layer and the dynamical events in this layer can be mimicked by choosing a suitable driver at the boundary. This approach has been quite successful since it allows us to study the large scale events such as CMEs (Coronal Mass Ejection), flares etc. in a very simple and elegant framework without unduly worrying about the microphysics of the {\it cold}, collisional region. Only difficulty with this approach is that it requires the presence of the fully ionized medium which does not exist over several pressure scale height \citep{VAL81}. 

The ideal MHD description of the cold solar plasma is often justified by using the text book argument based on the Reynolds number, $Re$ (which is the ratio of the fluid advection of the magnetic field to the field drift due to magnetic diffusion). It is argued that since $Re \gg 1$ for typical solar parameters, the magnetic diffusion can be neglected (except in the thin resonance layers) in the induction equation. However, even in the absence of diffusion, the plasma is still weakly ionized in the foot-point region of the flux tubes and thus the inertia of the fluid is due mainly to the neutral hydrogen. Therefore, the use of the ideal MHD of fully ionized plasma in the $Re \gg 1$ limit can not model the dynamics of the lower solar atmosphere.  

The partially ionized plasma can be described by a set of equations that is structurally similar to the {\it fully ionized} ideal MHD equations \citep{C57, BR65, PK96, P08a}, except now the non—-ideal MHD effects causes finite drift of the  magnetic field through the medium.  However, in the absence of current parallel to the magnetic field, the magnetic flux is frozen in the fluid notwithstanding its non—-idealness \citep{PW12}.\footnote{It should be noted here that in the absence of currents parallel to the magnetic field, the non—-zero Ohm diffusion does not affect the flux freezing–-a fact often missed in the MHD textbooks.} In the absence of field drift through the matter, i.e. in the absence of non—-ideal MHD effects, the set of equations in both the fully and partially ionized plasmas are formally identical except that the inertia of the fluid is carried by both the neutral and charged fluids. This, as we shall see has non-—trivial implications on the validity of the ideal—-MHD like description of the weakly ionized plasmas. Therefore, brushing aside the non—-ideal nature of the solar atmosphere in the long wavelength limit may not be entirely justified. 

Study of the magnetohydrodynamic (MHD) waves is thought to play an important role in the space and laboratory plasmas. The investigation of the \alf wave is important to the diverse physical settings. For example, \alf waves are possible source of turbulence in the interstellar medium; they are probably responsible for the heating of the solar corona; these waves are utilized as a source of fusion plasma heating.  Since \alf waves have non—-zero vorticity \citep{G12} they may transport angular momentum across the flux tube. 

By \alf waves, we often imply waves in the fully ionized, ideal MHD fluid where the ion inertia balances the field deformation. Frequency of these waves is often too high and thus may not survive the collisional damping in a partially ionized medium \citep{T62, KP69, U98, KR03, V07, V08, M11}.  However, partially ionized, collision dominated medium also supports low frequency \alf wave which is caused by the balance between the bulk fluid inertia and the deformation of the magnetic field. In the vanishing plasma inertia limit, this means that the inertia of the wave is solely due to neutral particles. The collisional momentum exchange plays the crucial role in transferring the magnetic stresses to the neutrals. Therefore, in the low frequency (in comparison with the neutral--ion collision frequency) limit, \alf wave propagates with very little damping in the magnetized medium \citep{KP69, P08a, P08b}.   

Owing to the presence of the magnetic field, density inhomogeneity and gravity, partially ionized medium in space is often highly structured. The known example includes the discs around the star and planet forming regions, layered structure of the solar atmosphere, Earth's ionosphere. The structured magnetized medium can support hydromagnetic waves whose behaviour at the surface boundary is considerably different from the bulk fluid \citep{G09}. At the interface between the structured layers, the amplitude of the surface wave decays exponentially across the surface while remaining constant along it. The investigation of these waves in the layered structures has been the subject of numerous studies in the past. The \alf surface wave can possibly play dominant role in heating the solar corona \citep{I78, G94, P97a, P97b, P98, A09}. The resonant absorption of the \alf waves in the inhomogeneous plasma layer has been suggested as a means of driving non—-Ohmic plasma heating in the toroidal devices \citep{T73, C74, K77, M92, P95}.  

Driven by the desire to understand the various observed wavelike features, hydromagnetic surface waves in the solar atmosphere in the framework of  ideal and Hall MHD have been discussed over the past several decades \citep{P74, R78, W79, R79, E82, S82, C85, C86, G94, Z96, R06, Z09, G09, G12}. Present work will primarily investigate the nature of the wave propagation in the partially ionized medium and discuss the results in the context of solar atmosphere, where various modes have possibly been identified in the recent past [cf.~Goossens et al. (2012) and references therein]. Investigation of the surface wave in the cylindrical filaments \citep{S09, S13} suggests the important role of the ambipolar diffusion in damping the short wavelength fluctuations. We build upon the past studies of the surface waves in the ideal and Hall MHD.  

This paper is organized in the following fashion. The basic model is briefly described in Sec.~2. In section 2.1 we discuss the validity of the ideal MHD description in the partially ionized medium before discussing the linear dispersion relation in some detail in subsection 2.2. In sec.~3, various boundary conditions are discussed and the dispersion relation is given for both sausage and kink modes. In sec.~4 discussion of the results and a brief summary is presented and future direction is indicated.  
\section{Basic model}

Although, the basic set of dynamical equations, which in the limiting case gives fully ionized, ideal MHD and weakly ionized, zero plasma inertia description as the two distinct limits, was formulated more than 50 years ago \citep{C57,BR65}, it was only recently that the spatial and temporal dynamical scales over which magnetic diffusion operates were clarified \citep{P06, P08a}. Thus, we shall make use of the MHD—-like formulation of the partially ionized solar atmosphere given by \cite{P08a}. 

We shall use the single fluid description of the partially ionized medium to investigate the properties of the  low frequency surface waves in an incompressible plasma slab. The general set of equations for the bulk fluid is 
\bq
\delt \rho + \grad \left(\cdot \rho\,\v \right)= 0\,.
\label{eq:cont}
\eq
Here $\v$ is the bulk velocity, $ \v = (\rho_i\,\vi + \rho_n\,\vn)/\rho$. The bulk mass density $\rho = \rho_i + \rho_n$ where $\rho_{i\,, n} = m_{i\,,n}\,n_{i\,,n}$ is the ion (neutral) mass and number densities respectively. Here $ m_{i\,,n}$ and $n_{i\,,n}$ are the mass and the number densities of the ion and neutral respectively.

The momentum equation is 
\bq
\rho\,\frac{d\v}{dt}=  - \nabla\,P + \frac{\J\cross\B}{c}\,,
\label{eq:meq}
\eq
where $\J = e\,n_e\,\left(\vi - \ve\right)$ is the current density, $\B$ is the magnetic field, $c$ is the speed of light, $e$ is the electronic charge and $P = P_e + P_i + P_n$ is the total pressure.
The induction equation can be explicitly written in terms of fluid ($\v$) and field ($\vB$) velocities as \citep{PW12}
\bq
\delt \B = \curl\left[
\left(\v + \v_B\right)\cross\B - \frac {4\,\pi\,\eta_O}{c}\,\Jpa\right]\,,
\label{eq:indA}
\eq
where the magnetic drift velocity ($\vB$) is defined as 
\bq
\vB = \eta_P\,\frac{\left(\grad\cross\B\right) \cross\hB}{B} -– 
\eta_H\,\frac{\left(\grad\cross\B\right)_{\perp} }{B}\,, 
\label{eq:md0}
\eq
with $\hB = \B /B$, $\Jpa = (\J\cdot \hB)\,\hB$ and $\eta_P = \eta_A + \eta$. The Ohm ($\eta$), ambipolar ($\eta_A$) and Hall ($\etaH$) diffusivities are given as
\bq \eta =
\frac{c^2}{4\,\pi\sigma}\,\,, 
\eta_A = \left(\frac{\rho_n}{\rho_i}\right)\,\frac{v_A^2}{\nu_{n\,i}}\,, \mbox{and}\, \etaH = \frac{v_A^2}{\omega_H}\,.
\label{eq:diffu}
 \eq
Here $\sigma = c\,e\,n_i \,\left( \beta_e + \beta_i \right) / B$ is the parallel conductivity defined in terms of plasma Hall parameter 
\bq
\beta_j = \frac{\omega_{c\,j}} {\nu_{j\,n}}\,, 
\eq
which is a ratio between the plasma—-cyclotron ($\omega_{cj} = e_j\,B / m_i\,c$ with $e_j = \pm e$) and plasma-—neutral ($\nu_{jn} = \rho_n\,\nu_{nj} / \rho_j$) collision frequencies. The \alf velocity   $v_A = B / \sqrt{4\,\pi\,\rho}$, and the Hall frequency $\omega_H $ \citep{P08a} 
\bq
\omega_H = \frac{\rho_i}{\rho}\,\omega_{ci}\,.
\label{eq:hfy}
\eq
The Hall scale $L_H$ in the partially ionized medium is a function of the fractional ionization $X_e = n_e / n_n$ and can be written as \citep{P08a}

\bq
L_H \cong X_e^{-1/2}\,\delta_i\,,          
\label{eq:HSL}
\eq 
where $\delta_i = v_{A\,i} / \omega_{ci}$ is the ion--inertial scale. Note that 
here $ v_{A\,i} = B / \sqrt{4\,\pi\,\rho_i}$ is the \alf speed in the ion fluid. 

Recall that the various diffusivities depend on the magnetization of the medium \citep{PW13}. For example, when the electrons are well coupled or partially coupled to the field and the ions are partially or completely decoupled from the field, Hall drift of the field dominates over other diffusive processes. When the magnetic field can be regarded as frozen in the plasma and drifts with it through the neutrals (applicable to the relatively low-density, high ionization-fraction regions) the ambipolar diffusion is the dominant diffusion in the medium. When the neutrals stop the ionized particle from drifting with the field, Ohm becomes the dominant diffusion mechanism.  Therefore, we should expect that the various diffusive scales in the solar atmosphere will mainly be dependent on the fractional ionization ($X_e$) and  plasma magnetization ($\beta_j$).

The ambipolar $L_A$ and Ohm $L_O$ scales can be expressed in terms of Hall scale $L_H$ via the ion ($\beta_i$) and electron ($\beta_e$) Hall parameters \citep{PW13}
\bq
L_A = \beta_i\,L_H\,,\quad L_O = \beta_e^{-1}\,L_H\,.
\eq
Although the Hall scale is independent of the ambient magnetic field strength, both the ambipolar and Ohm scale depends on the magnetization of the medium. Thus in order to calculate the changes in the various diffusive scales with altitude, we shall adopt a flux tube model from \cite{PW13} and assume the presence of a kG field at the footpoint. For the ion and neutral density profiles in the solar photosphere-—chromosphere, we use the model C from \cite{VAL81}. 
\begin{figure}
\includegraphics[scale=0.30]{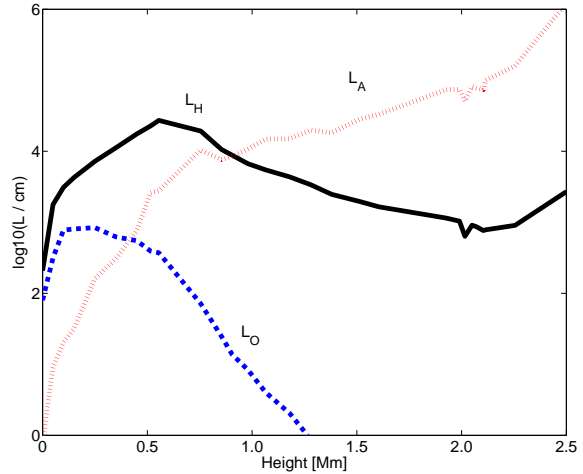}
\caption{\large{The variation of the Hall ($L_H$), ambipolar ($L_A$) and Ohm ($L_O$) scales in the photosphere—-chromosphere region is shown in the above figure. We have assumed $m_i = 30\,m_p$ where $m_p$ is the proton mass. A kG field is assumed at the footpoint.}}
\label{fig:fig1}
\end{figure}
It is clear from the above figure that for a kG magnetic field at the base of the flux tube, Hall is the dominant diffusion mechanism in the photosphere and lower chromosphere. Since the Hall scale is typically of the order of $\sim \mbox{km}$, the Hall drift of the field becomes important over few $\mbox{kms}$. Ambipolar diffusion scale becomes comparable to the Hall in the lower chromosphere ($\sim \mbox{Mm}$) before dominating it with increasing height. Clearly wave propagation in a thin flux tube ($\lesssim 100\,\mbox{km}$ diameter) will be affected by both the Hall and ambipolar diffusion.  For a weaker field, above discussion is still valid except now the Ohm dissipation length becomes comparable to the Hall scale near the footpoint.

Above set of Eqns.~(\ref{eq:cont}), (\ref{eq:meq}) and (\ref{eq:indA}) can be closed by assuming $P = P (\rho)$. However, we shall consider an incompressible fluid and thus, the pressure term will drop out of the linearized equations. Furthermore, we shall assume that the magnetic flux is frozen in the medium moving with $\v + \vB$, i.e. neglect the last term in the induction equation. This implies that either 
the field aligned parallel current $\Jpa = 0$ or, the Ohm diffusion is unimportant. The analysis of the three—-dimensional vector currents in a sunspot from the photosphere to the chromosphere suggest that the currents in general are not aligned with the field \citep{S05}. However, since with the increasing height, the current becomes predominantly parallel to the field in the chromosphere, the neglect of $\eta\,\Jpa$ can be justified since Hall and ambipolar becomes the dominant diffusion mechanism with increasing altitude \citep{PW13}. Thus we shall neglect $\eta\,\Jpa$ in the induction equation. 
\subsection{Ideal MHD of the partially ionized plasmas}
In the absence of magnetic field drift, $\vB = 0$, i.e. setting $\eta = \eta_P = \eta_H = 0$, the set of Eqns.~(\ref{eq:cont}), (\ref{eq:meq}) and (\ref{eq:indA}) becomes identical to the ideal MHD equations, except now the plasma is partially ionized.  It is well known that the ion Larmour radius provides an implicit scale in the ideal MHD theory \citep{F82} and thus the validity of ideal MHD requires that the characteristic length scale should be larger than the Larmor radius. How does this requirement affect the ideal MHD of partially ionized plasmas? In order to answer this question, we write the electric field in the rest frame of the neutrals, which in the absence of diffusion becomes [Eq.~(22), \cite{P08a}]   
\bq
 \E + \frac{\v\cross\B}{c} = \frac{\J\cross\B}{c\,e\,n_e} - \frac{\grad{P_e}}{c\,e\,n_e}\,.
\label{eq:ef1}
\eq 
Clearly, the ideal MHD limit requires that the right hand side terms are negligibly small. Thus imposing 
\bq
|\frac{\J\cross\B / e\,n_e}{\v\cross\B}| \ll 1\,,
\eq
leads to $\J / e\,n_e\,\v \ll 1$. Note that 
\bq
\frac{\J}{e\,n_e\,\v} \sim \frac{c\,P}{L\,e\,n_e\,v\,B}\,,
\eq  
where L is the characteristic scale size over which ideal MHD is valid. Assuming $v \sim c_s  = \sqrt{ k_B\,T / m_n}\,$ and, $P = c_s^2\,\rho_n$ we get
\bq
\frac{\J}{e\,n_e\,\v} \sim X_e^{-1}\,\left(\frac{r_L}{L}\right)\,.
\eq 
Here $r_L = c_s / \omega_{ci}$ is the Larmor radius, which is implicit scale of the ideal MHD in the fully ionized plasma. In the ideal MHD description of the partially ionized plasma, the implicit scale becomes 
\bq
R_L = X_e^{-1}\,r_L\,. 
\label{eq:mlR}
\eq 
Therefore, the ideal MHD like description is valid in the partially ionized plasmas if the characteristic scale length $L$ is larger than the modified Larmor radius $R_L$.  
To summarize, although the ideal MHD equations for both the fully ionized and partially ionized plasmas looks similar, the scale of their validity is quite different. It is clear that in a weakly ionized medium ($X_e \ll 1$) only long wavelength (and hence low frequency) fluctuations can be described by the MHD like equations. The short wavelength fluctuations, which otherwise would have propagated in a fully ionized medium ($X_e \gg 1$) will undergo damping.  
This conclusion could also have been anticipated on the ground that the validity of the single fluid ideal MHD like description of the partially ionized plasma requires 
\bq
\omega \lesssim X_e^{-1/2}\,\nu_{in}\,,
\label{eq:sff}
\eq 
where we have used $\rho_n\,\nu_{ni} = \rho_i\,\nu_{in}$, and made following simplifying assumptions $1 + \rho_n \,\beta_e/ \rho \approx \rho_n \,\beta_e/ \rho$ and $m_i = m_n$ in the Eq.~(10) of \cite{P08a}. 
\begin{figure}
\includegraphics[scale=0.30]{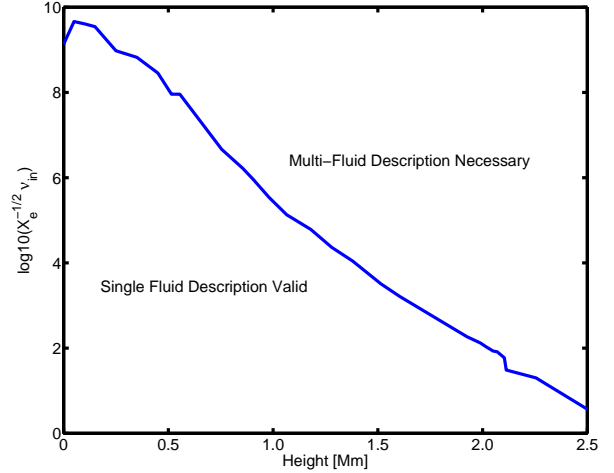}
\caption{\large{The region where single fluid formulation of the multi—component plasma is valid is shown in above figure for the parameters pertaining to the solar photosphere—-chromosphere.}}
\label{fig:fig2}
\end{figure}

Making use of the Table~1 from \citep{PW13}, the right hand side of the above Eq.~(\ref{eq:sff}) is plotted in Fig.~(\ref{fig:fig2}). We see from the figure that in the photosphere ($\lesssim 0.5\,\mbox{Mm}$), the low frequency / long wavelength ($\sim 10^6-—10^8\,\mbox{HZ}$) fluctuations in the medium can be investigated in the framework of the single fluid formulation. With increasing altitude, the single fluid formulation remains valid only for the smaller frequencies. Therefore, at the chromosphere—-corona transition region ($\gtrsim  2.5\,\mbox{Mm}$), very low frequency behaviour of the plasma is amenable to the single fluid description. All in all, low frequency long wavelength behaviour of the photosphere—-chromosphere plasma is amenable to the single fluid MHD--like description [see also \cite{SK12}]. Therefore, notwithstanding the structural similarity of the fully and partially ionized plasmas, we shall keep above difference in mind while discussing the ideal MHD limit of the generalized surface wave dispersion relation.  
\subsection{Dispersion Relation}

The magnetic field in the photosphere—-lower chromosphere is structured in the form of flux tubes and constitute the well known magnetic network. The tubes are predominantly vertical and in the pressure equilibrium with the outside medium. Often a simplified model of the flux tube to investigate the wave properties of such a medium is approximated by a plasma slab with piecewise constant density \citep{E82}.  Thus we shall consider a partially ionized, incompressible, magnetized slab of thickness $2\,x_0$ permeated by the uniform vertical magnetic field $\B = B \,\hat{\bmath{z}}$. The thickness of the slab represents the diameter of the flux tube.   

The equilibrium state is described by the following piecewise continuous density profile,
\bq
\rho(x) = \left\{
\begin{array}{ll}
\rhon & \mbox{if $|x|  \leq x_0$};\\
\rhox & \mbox{if $|x| > x_0$}\,, 
\end{array}
\right.
\eq 
This density discontinuity fundamentally changes the behaviour of the \alfc vorticity propagation in the medium. Whereas in an infinite uniform medium, the \alfc vorticity in nonzero in the entire volume, the density jump confines the vorticity to the surface layer $x = x_0$ only \citep{G12}. 

The linear perturbation around the equilibrium is described by the following equations
\begin{eqnarray}
\nabla\cdot\dv& = &0\,,
\nonumber \\
\delt\dv & = & - \nabla\left( \frac{\B\cdot\dB}{4\,\pi\,\rho}\right) + \frac{\left(\B\cdot\nabla\right)\dB}{4\,\pi\,\rho}\,, 
\nonumber \\
\delt\dB & = & \left(\B\cdot\nabla\right)\left(\dv  + \delta \vB\right)\,, 
\label{eq:Li}
\end{eqnarray}
along with the $\nabla\cdot\dB = 0$. 
Fourier analyzing the perturbed quantities as $exp{\left(-i\,\omega\,t + i\,k\,z\right)}$, and defining 
${\bf{L}} = \frac{d^2}{dx^2} -– k^2\,,\, \mbox{and}\, \omega_A^2 = k^2\,v_A^2$, the momentum and induction equations can be written in the following form respectively. 
\bq
\left(
\begin{array}{cc} \omega   &  0\\
                   0       & \omega
  \end{array}
\right)\,\dv
 = v_A^2
\left(
\begin{array}{cc} \frac{1}{k}{\bf{L}}   &  0\\
                   0       & - k
  \end{array}
\right)\,\frac{\dB}{B}\,,
\label{eq:lin1}
\eq
and,
\begin{eqnarray}
\left(
\begin{array}{cc} \omega –- i\,\eta_P\,{\bf{L}}   &  0\\
                   i\,\eta_{H}\,{\bf{L}} & \omega + i\,k^2\,\eta_{P}
  \end{array}
\right)\,\frac{\dB}{B}
\nonumber\\
 = k\left(
\begin{array}{cc} - 1   &  i\,\frac{\eta_H}{v_A^2}\,\omega\\\
                   0       & - 1
  \end{array}
\right)\,\dv\,.
\label{eq:lin2}
\end{eqnarray}
Eqs.~(\ref{eq:lin1}) and (\ref{eq:lin2}) can be reduced to the following coupled differential equation  
\begin{eqnarray}
\left(\frac{d^2}{dx^2} –- k^2\right)\dvx
+ F_1\left(\frac{d^2}{dx^2} –- k^2\right)\dvy = 0\,,
\label{eq:xx1}
\\
\left(\frac{d^2}{dx^2} –- k^2\right)\dvx –- i\,F\,\dvy = 0\,,
\label{eq:xx2}
\end{eqnarray}
where
\begin{eqnarray}
F = \left(\frac{\omega}{\eta_H}\right) \left[ 1 -– a + i\,\frac{k^2\,\eta_P}{\omega}\right]\,,
\nonumber\\
F_1 = \frac{i\,\eta_P}{\eta_H} \frac{1}{\left(1 –- a\right)}
\left[ 1 –- a + i\,\frac{k^2\,\eta_P}{\omega}\left(1 + \frac{\eta_H^2}{\eta_P^2}\right) \right]\,, 
\label{eq:FDF}
\end{eqnarray}
and $a = \oma^2 / \omega^2$. It is clear from Eq.~(\ref{eq:FDF}) that for $\eta_P = 0$, $F \sim  1 / \eta_H$ and $F_1 \sim \eta_H$. Therefore, the coupling between $\dvx$ and $\dvy$ in Eq.~(\ref{eq:xx1}) and Eq.~(\ref{eq:xx2}) is entirely due to the Hall effect. 

In a homogeneous medium we may Fourier analyze the $x$--dependence as $\exp{\left(i\,n\,x\right)}$ and thus Eqns.~(\ref{eq:xx1})–-(\ref{eq:xx2}) reduces to the following dispersion relation
\begin{eqnarray}
\left(\omega^2 - \omega_A^2\right)\left[\left(\omega^2 - \omega_A^2\right) + i\,k^2\,\eta_P\,\omega\right] + i\chi^2\,\eta_P\,\omega  
\nonumber\\
\times \left[\left(\omega^2 - \omega_A^2\right) + i\,k^2\,\eta_P\,\omega\left( 1 + \frac{\eta_H^2}{\eta_P^2}\right)\right] = 0\,,
\label{eq:drn}
\end{eqnarray}
where $\chi^2 = n^2 + k^2$. In the absence of Pedersen diffusion, Eq.~(\ref{eq:drn}) reduces to 
\bq
\omega^2 = \omega_A^2 \pm \left(\frac{\chi}{k}\right)\,k^2\,\eta_H\,\omega\,, 
\eq
which for the short wavelength ($\omega_H \ll \omega_A$) fluctuations, describes the whistler waves in the high frequency ($\omega_A \ll \omega$) limit  
\bq
\omega \cong  \left(1 + \frac{n^2}{k^2}\right)^{1/2}\,k^2\,\eta_H\,,
\eq
and, electrostatic ($\curl \dE \approx 0$) ion–-cyclotron wave in the low frequency $\omega \ll \omega_A$ limit,   
\bq
\omega \cong \left(1 + \frac{n^2}{k^2}\right)^{-1/2}\,\omega_H\,.
\eq
 In the long wavelength limit, i.e. $\omega_A \ll \omega_H$, we recover usual \alf wave $\omega^2 = \omega_A^2$ (cf.~ Pandey and Wardle (2008)).

In the absence of Hall, $\eta_H = 0$, when only Pedersen diffusion is present,
the dispersion relation, Eq.~(\ref{eq:drn}) reduces to
\begin{eqnarray}
\left[\left(\omega^2 - \omega_A^2\right) + i\,k^2\,\eta_P\,\omega\right] \left[\left(\omega^2 - \omega_A^2\right) + i\,\chi^2\,\eta_P\,\omega\right] = 0\,,
\label{eq:dra}
\end{eqnarray}
which gives
\bq
\omega = \pm \left[ \omega_A^2 - \frac{1}{2}\left(\chi^2\,\eta_P\right)^2\right]^{1/2} –-
 i \frac{\chi^2\,\eta_P}{2}\,.
\label{eq:alf}
\eq 
Eq.~(\ref{eq:alf}) is identical to Eq.~(C6) of \cite{KP69}. Recall that Eq.~(C6) of \cite{KP69} is derived in the low frequency limit, i.e. $\omega \lesssim \omega_A \ll \nu_{ni}$. 

Seeking the solution of Eqns.~ (\ref{eq:xx1})--(\ref{eq:xx2}) as
\begin{eqnarray}
\dvx & = & f \left[ \exp^{-\alpha\, x} \mp \exp^{\alpha\, x} \right]\nonumber\\ 
\dvy & = & i\,h\, \left[ \exp^{-\alpha\, x} \mp \exp^{\alpha\, x} \right]\,, 
\end{eqnarray} 
We obtain the following set of equations 
\begin{eqnarray}
\left(\alpha^2 -– k^2\right)\,\left[ f + h\,F_1\right] & = & 0\,,\nonumber\\
\left(\alpha^2 -– k^2\right)\,f + h\,F & = & 0\,,
\end{eqnarray}
which yields
\begin{eqnarray}
\alpha_1 & = & k\,,
\nonumber\\
\alpha_2 & = & k\,\left[ 1 + \frac{F}{k^2\,F_1} \right]^{1/2} \equiv m\,.
 \label{eq:rta}
\end{eqnarray}
This means that there are two pairs of attenuation coefficients, $\left(k\,,m_{in}\right)$ inside and $\left(k\,,m_{ex}\right)$ outside the slab. Note that in general, $\alpha_2$ is complex since
\begin{eqnarray}
\mbox{Re}\left[\frac{F}{F_1}\right] & = & \frac{k^2\,\eta_{\perp}^2 / \eta_H^2}
{\frac{\left(k^2\,\eta_{\perp}^2\right)^2}{\eta_H^2\,\omega^2\,\left(1-a\right)^2} + \eta_P^2 / \eta_H^2}
\,,\nonumber \\
\mbox{Im}\left[\frac{F}{F_1}\right] & = & \left( \frac{k^2\,\eta_P}{\eta_H}\right)\,
\left[\frac{
\frac{k^2\,\eta_{\perp}^2}{\eta_H\,\omega\,\left(1-a\right)}
- \frac{\omega\,\left(1-a\right)}{k^2\,\eta_H}
}
{\frac{\left(k^2\,\eta_{\perp}^2\right)^2}{\eta_H^2\,\omega^2\,\left(1-a\right)^2} + \eta_P^2 / \eta_H^2}
\right]\,,
\end{eqnarray}
 where $\eta_{\perp}^2 = \eta_P^2 + \eta_H^2$.  It is clear from the above expressions that when $\eta_P = 0$, $\mbox{Im}[ F / F_1] = 0$ and the wavenumber is real. This is not surprising since there is no wave attenuation in the Hall plasma.  

Guided by the fact that the planer or cylindrical waveguides can support kink and sausage modes, we formally choose the general solution of the $\dvx$ and $\dvy$ as a superposition of these waves. 

For the sausage wave, inside the slab ($|x| \leq x_0$)
\begin{eqnarray}
\dvx (x) & = & f_1\,\frac{\sinh(k\,x)}{ \sinh(k\,x_0)} + f_2\,\frac{\sinh(m_{in}\,x)}{ \sinh(m_{in}\,x_0)}\,,\\
\dvy (x)& = & i\,f_1\,G_{in1}\,\frac{\sinh(k\,x)}
{\sinh(k\,x_0)} + i\,f_2\,G_{in2}\,\frac{\sinh(m_{in}\,x)}{ \sinh(m_{in}\,x_0)}\,,\nonumber\\
{}
\label{eq:sw1}
\end{eqnarray} 
where 

\bq
G_{in1\,,2} = - \eta_{Hin}\,\frac{\left(\alpha_{in1\,,2}^2 -– k^2\right)}{\left[ \left(1 –- a_{in}\right)\,\omega + i\,k^2\,\eta_{Ain}\right]}\,.
\eq
Note that $G_{in1} = 0$. For the kink surface–-waves, identical expression for the perturbed velocities can be given by replacing $\sinh$ by $\cosh$. The solution outside the plasma layer is
\bq
%\begin{eqnarray}
\dvx = \left\{ \begin{array}{ll}
s_1\,e^{- k\, \left(x -– x_0\right)} + s_2\,e^{-m_{ex}\, \left(x -– x_0\right)} & \mbox{if $x > x_0$}\,,\\
\beta_1\,e^{k\, \left(x + x_0\right)} + \beta_2\,e^{m_{ex}\, \left(x + x_0\right)} & \mbox{if $x < - x_0$}\,, 
\end{array}
\right.
%\end{eqnarray} 
\eq
and 
%\begin{eqnarray}
\bq
\dvy = \left\{ \begin{array}{ll}
i\,s_1\,G_{ex1}\,e^{- k\,\left(x -– x_0\right)} + i\,s_2\, G_{ex2}\,e^{-m_{ex}\,\left(x -– x_0\right)} & 
\\
\mbox{if $x > x_0$}\,,\\
%\,,\nonumber\\
 %x > x_0\nonumber\\ 
%\nonumber\\ 
%\dvy = 
i\,\beta_1\,G_{ex1}\,e^{k\,\left(x + x_0\right)} + i\,\beta_2\, G_{ex2}\,e^{m_{ex}\, \left(x + x_0\right)} & \\
\mbox{if $x < - x_0$}\,.
 %\,,\nonumber\\ x < - x_0\,. 
%\end{eqnarray} 
\end{array}
\right.
\eq
Here
\bq
G_{ex1\,,2} = - \eta_{Hex}\,\frac{\left(\alpha _{ex1\,,2}^2 -– k^2\right)}{\left[ \left(1 –- a_{ex}\right)\,\omega + i\,k^2\,\eta_{Aex}\right]}\,.
\eq
and $G_{ex1} = 0$. 
The knowledge of $\dvx$ and $\dvy$ allows us to calculate the perturbed total pressure
\begin{eqnarray}
\delta p_T = \frac{\B\cdot\dB}{4\,\pi} = i\,\frac{\rho\,\va^2}{\omega} \left[\left(\frac{1 - a}{a}  \right)\frac{d\dvx}{dx} + 
\right.
\nonumber\\
\left.
\left\{\frac{\eta_P}{\eta_H}\left(\frac{1 - a}{a} + i\,\frac{\eta_P}{\eta_H}\left(\frac{\omega}{\omega_H}\right)\right) + i\,\left(\frac{\omega}{\omega_H}\right) \right\}\frac{d\dvy}{dx}
\right]\,.
\end{eqnarray}
The electric field components $\dEx$ and $\dEy$ which will be required for the boundary conditions, can be derived from the generalized Ohm${}^\prime$s law
\bq
c\,\dE = - \left(\dv + \delta \vB\right)\cross\B\,, 
\eq
which yeilds
\bq
\frac{c\,\dEx}{B} = - \frac{1}{a} \,\dvy\,,
\eq
and
\begin{eqnarray}
\frac{c\,\dEy}{B} = \dvx - \left[\frac{\eta_P}{\eta_H}\left(\frac{1 - a}{a} + i\,\frac{\eta_P}{\eta_H}\left(\frac{\omega}{\omega_H}\right)\right)\right. \nonumber\\
+ \left. i\,\left(\frac{\omega}{\omega_H}\right) \right]\,\dvy \,.
\end{eqnarray}
\section{Surface Waves}
We need four boundary conditions across $x = x_0$ in order to derive the dispersion relation \citep{Z96}. Thus, the first boundary condition, the continuity of the total pressure across the slab $[\delta p_T] = 0$ gives following relation
\begin{eqnarray}
X_{in1}\,f_1\,k\,\tanh\left(k\,x_0\right) + X_{in2}\, f_2\,m_{in}\,\tanh\left(m_{in}\,x_0\right) \nonumber\\
= - \left( 
X_{ex1}\, s_1\,k + X_{ex2}\, s_2\,m_{ex} 
\right)
\label{eq:bc1}
\end{eqnarray} 
where $X_{j} = Y_j + i\,Q_j\,G_{j}$ with 
\bq
Y = \left(\frac{ 1 -– a}{a}\right)\,,  
\eq
and
\bq
Q = \frac{\eta_{A}}{\eta_{H}}\left( Y + i\,\frac{\eta_{A}}{\eta_{H}}\left(\frac{\omega}{\omega_H}\right)\right)  + i\,\left(\frac{\omega}{\omega_{H}}\right)\,. 
\eq
The second boundary condition is derived by integrating $\dvx$ across the boundary which yields $f_1 + f_2 = s_1 + s_2$. Third boundary condition is derived by integrating induction equation. For this purpose, we need to write the induction equation in the following conservative form
\bq
\delt\dB + \div \delta\U = 0\,,
\label{eq:ind1}
\eq
where
\bq
\delta \U = \left(\delta\v + \delta \vB\right)\,\B - \B\,\left(\delta\v + \delta \vB\right)\,, 
\eq
and, integrating Eq.~(\ref{eq:ind1}) across the surface layer gives $[\delta \U] = 0$. As the fourth boundary condition, we impose the continuity of the electric displacement across the surface \citep{Z96} $\left[\delta D_x \right] = 0$  where $\delta D_x \cong K_{xx}\, \dEx + K_{xy}\, \dEy$ with
\bq
K_{xx} \approx \frac{c^2}{\va^2}\,\,,K_{xy} \approx i\,\frac{c^2}{\va^2}\,\left(\frac{\omega}{\omega_H}\right)\,.
\eq
By imposing above boundary conditions and defining $S_1 = Q_{in} / Q_{ex}$, we arrive at the following dispersion relation
\begin{eqnarray}
\left(\frac{\omega^2}{\oma_{in}^2} - 1\right)\,\,\left[T_{in1} + T_{in2} + i\,\frac{\eta_P}{\eta_H}\,G_{in2}T_{in2}\right] 
\nonumber\\
- S_{1}\left(\frac{\omega^2}{\oma_{ex}^2} - 1\right) 
 \left[\frac{\left(k\,T_{ex} - m_{ex}\,G_{in2} \right)}{G_{ex2}} 
+ i\, m_{ex}\,\frac{\eta_P}{\eta_H}\,G_{in2}\right] 
\nonumber\\
- \left(\frac{\omega}{\omega_H}\right)\, \left(1 + \frac{\eta_P^2}{\eta_H^2}\right)\,G{_{in2}}\,\left[
T_{in2}+ m_{ex}\,S_{1}
\right]  = 0 \,,
\label{eq:dr1}
\end{eqnarray}
where 
\begin{eqnarray}
T_{in1} = - k\,\left(1 -– C\,G_{in2}\,S_1\right) 
\left\{     \begin{array}{l}
\tanh \left(m_{in}\,x_0\right)\,,\\
\coth \left(m_{in}\,x_0\right)\,.
\end{array}
\right.
\nonumber\\
T_{in2} =  m_{in}\, 
\left\{     \begin{array}{l}
\tanh \left(m_{in}\,x_0\right)\,,\\
\coth \left(m_{in}\,x_0\right)\,.
\end{array}
\right.
\quad
T_{ex} = G_{in2}\left( 1 -– C\,G_{ex2}\right)\,,
%\nonumber\\
%N =  - m_{e}\,G_{e2}\,T_{e1}\,,
\end{eqnarray}
and
\bq
%S3 = \left(1 -– C\, G_{i2}\,S_1\right) \,,
%\quad
C =\left(\frac{\omega}{\omega_H}\right)^{-1}\, 
\frac{A_{2} - A_{1}}{\left(\rhon / \rhox - 1 \right)}\,.
\eq
Here
\bq
A_1 = \frac{\rhon\, Q_{ex}}{\rhox\, Q_{in}}\left[c_{1} - \left(\frac{\omega}{\omega_{Ain}}\right)^2\, c_{2}\right]\,,
\eq
and
\bq
A_2 = \left[c_{1} - \left(\frac{\omega}{\omega_{Aex}}\right)^2\, c_{2}\right]\,,
\eq
with
\begin{eqnarray}
c_1 = \left(\frac{\omega}{\omega_H}\right)^2
\left(1 + \frac{\eta_P^2}{\eta_H^2}\right)
+ i \,\frac{\eta_P}{\eta_H} \left(\frac{\omega}{\omega_H}\right)
\,,\nonumber\\
c_2 = 1 + i \,\frac{\eta_P}{\eta_H}\,\left(\frac{\omega}{\omega_H}\right)\,.
\end{eqnarray}
We note that the ratio of the Hall and ambipolar diffusivities is assumed constant inside and outside the tube implying that the ratio of the Hall ($\omega_H$) to the neutral –- ion collision ($\nu_{ni}$) frequencies is constant. This is a plausible assumption.   

In the absence of Hall diffusion ($\eta_H = 0$) Eq.~(\ref{eq:xx1}) -- (\ref{eq:xx2}) reduces to the following form
\begin{eqnarray}
\left[\frac{d^2}{dx^2} –- k^2 \left( \frac{\left(\omega^2 - \omega_A^2\right) + i\,k^2\,\eta_P\,\omega}{- \omega_A^2 + i\,k^2\,\eta_P\,\omega}\right)
\right]\dvx = 0\,,\nonumber\\
\left( \omega^2 - \omega_A^2 + i\,k^2\,\eta_P\,\omega\right)\dvy = 0\,. 
\label{eq:pamb}
\end{eqnarray}
We may infer from the $\dvy$ equation above that the wave damps at a rate give by Eq.~( \ref{eq:alf}). The $\dvx$ equation gives $\dvx = Q\,x + Const.$. Since the fluctuation decay time is of the order of ambipolar diffusion ($\sim 1 / k^2\,\eta_A$), the physical solution requires that the $const. = - Q\,x_0$ at the surface boundary. Thus, in the purely ambipolar regime, fluctuations will disappear at the surface boundary over $\sim 1 / k^2\,\eta_A$. 

The dispersion relation, Eq.~(\ref{eq:dr1}) is quite complicated and needs to be simplified to understand the role of diffusion on the surface waves. The general form of $m$, Eq.~(\ref{eq:rta}) implies that we have an implicit, transcendental equation for the phase velocity $V_P = \omega / k\,v_{Ain}$ against $k\,x_0$. There in no general prescription available to solve this dispersion relation. Therefore, we analyse it in various limiting cases. 

In order to gain the analytical insight on the role of ambipolar diffusion on the surface waves, we shall examine the dispersion Eq.~(\ref{eq:dr1}) in the long wavelength ($K \rightarrow 0$) limit assuming  $\eta_P / \eta_H \gg 1$. Defining $R = \rhox / \rhon$, $k x_0 \equiv K$, $H = L_H / x_0$, and normalized phase speed $V_{P} = \omega / \omega_{Ain}$, we may write 
\begin{eqnarray}
Q_{in} \approx \frac{\eta_P}{\eta_H}\,V_{P}^2\,, Q_{ex} \approx R\,Q_{in}\,,S_1 =  1/R\,,
\nonumber\\
A_{1, 2} \approx –- R_{q}\,c_1\, V_{P}^2\,,G_{in2} = G_{ex2} \approx i\,\frac{\eta_H}{\eta_P}\,,
\nonumber\\
\tanh(K) \approx K\,,
C\approx \frac{R\, V_P}{H\,K}\left( 1 + i\,D\,H\,K\right)\,,
\nonumber \\
T_{ex} \approx  \frac{R\,V_P}{D\,H\,K}
\,,T_{in1} \approx  0\,,
\end{eqnarray}
where $D = \eta_P / \eta_H\,, H = L_H / x_0$ and $R_{q} = 1\,,R$ for $q = in\,,ex$ respectively. Since $F \approx \omega / \eta_H$ and $F_1 \approx i\,D$, 
\bq
T_{in2} \approx \left(m_{in}\,x_0\right)^2 = \frac{-i\,V_P\,K}{D\,H}\,.
\eq
With the above approximations, the dispersion relation Eq.~(\ref{eq:dr1}) reduces to the following form
\bq
\left(\frac{\omega}{k\,v_{Ain}}\right) \approx \frac{-i\,\rhon}{\rhox}\,\left(\frac{\eta_P}{\eta_H}\right)^2\,,
\label{eq:drt}
\eq
suggesting that the long wavelength surface waves are strongly damped when Pedersen is the dominant diffusion in the plasma. Note that the damping rate also depends on the density ratio owing to the nature of the wave. 
  
In the Hall case, setting $\eta_P = 0$, the dispersion relation, Eq.~(\ref{eq:dr1}) reduces to the following form
\begin{eqnarray}
\left(\omega^2 - \omega_{Ain}^2\right) + \frac{\rhox}{\rhon}\left(\omega^2 - \omega_{Aex}^2\right) \tanh(k\,x_0) 
\nonumber\\
 - \left(\frac{\omega}{\omega_H}\right)^2\,\omega_{Ain}^2\,\left[1 + \frac{\rhon}{\rhox}\tanh(k\,x_0)\right] \left(1 + \frac{\rhon}{\rhox} \right)^{-1} = 0\,,
\label{eq:SHC}
\end{eqnarray}
which is similar in form to Eq.~(18) of \cite{Z96}, except now, the plasma is partially ionized and thus the \alf and Hall frequencies pertain to the partially ionized medium. Note that in a fully ionized medium Hall frequency reduces to the cyclotron frequency.  Above dispersion relation can be written as
\bq
V_{P}^2 = \frac{1 + \tanh{K}}
{1 + R\,\tanh{K} - \frac{H^2\,K^2\left(R + \tanh{K}\right)} 
{1 + R}}\,.
\label{eq:NHD}
\eq
It is clear from Eq.~(\ref{eq:NHD}) that in the absence of Hall, the phase speed 
is independent of the scale of the system. This is related to the generic nature of the ideal MHD like behaviour of the multi-–component system in the low frequency limit. The Hall term, which appears in the induction equation as an additional term, introduces a characteristic length scale $L_H$ to the otherwise scale--free equation. This is reflected in the denominator Eq.~(\ref{eq:NHD}). The $V_P$ is real only if 
\bq
k\,L_H \lesssim \sqrt{1 + R}\,\frac{\left(1 + R\,\tanh{K}\right)}{\left(R + \tanh{K}\right)}\,.
\label{eq:scd}
\eq      

In the absence of Hall, i.e. when $H = 0$, for a thin tube, Eq.~(\ref{eq:NHD}) reduces to 
\bq
V_{P}^2 \simeq 1 + (1 -– R)\,K\,,
\label{eq:ERD}
\eq
which is Eq.~(13) of \cite{E82}, except now the waves with wavelength shorter that $R_L$ [Eq.~(\ref{eq:mlR})] can not propagate in the medium. Furthermore, as noted above, the inertia of the fluid is due to the {\it bulk} fluid and not  due to the ion fluid. Therefore, although the dispersion relation, Eq.~(\ref{eq:ERD}) is formally similar to the Eq.~(13) of \cite{E82}, the nature of wave in the present case is different.

For a thick flux tube, when $K \gg 1$, Eq.~(\ref{eq:NHD}) becomes  
\bq
V_{P}^2 = \frac{2}
{\left(1 + R - H^2\,K^2\right)}\,.
\label{eq:NHD1}
\eq
which without the Hall, is $ V_P^2 = 2 / \left(1 + R\right)$, or, Eq.~(14) of \cite{E82}. As has been noted above, Hall effect introduces a scale in the otherwise scale free description of the bulk fluid and this leads to the above condition, Eq.~(\ref{eq:scd}) on the wavelength of the surface wave. Note that in a wide tube, when $k\,x_0 \gg 1$, Eq.~(\ref{eq:scd}) implies that $k\,L_H \lesssim \sqrt{1 + R}$, i.e. $L_H \lesssim x_0$, or, the width of the flux tube is on the order of the Hall scale. Since Hall scale shrinks to the ion–-inertial scale in a fully ionized medium, $L_H \lesssim x_0$ may be easily satisfied. However, in a weakly ionized medium, the Hall scale increases with the drop in the ionization [Eq.~(\ref{eq:HSL})] and thus it may not be always possible to satisfy the inequality $L_H \lesssim x_0$ implying that the propagation of the surface waves in a thick flux tube could be difficult in the Hall regime. 

\begin{figure}
\includegraphics[scale=0.30]{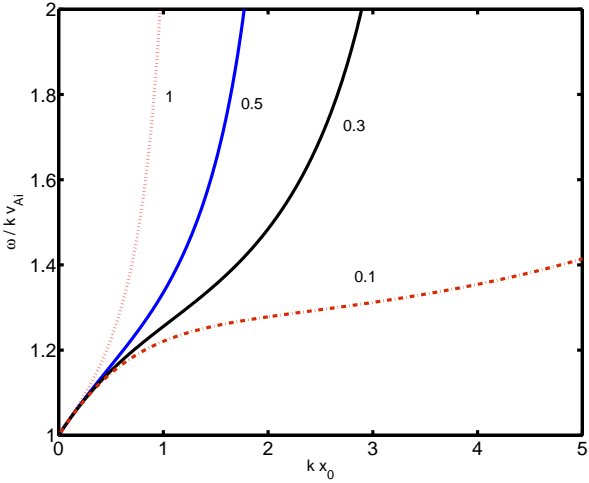}
\caption{\large{The normalized phase speed $V_P \equiv \omega / \omega_{Ain} $ of the sausage surface waves against $k\,x_0$ for $L_H / x_0 = 0.1, 0.3, 0.5$ and $ 1$ and $\rhox / \rhon = 0.25$ is shown in the figure.}}
\label{fig:Haf}
\end{figure}

We solve Eq.~(\ref{eq:SHC}) by assuming that the ratio of the densities inside and outside the tube $\rhox / \rhon = 0.25$. In Fig.~ (\ref{fig:Haf}) we plot $\omega / k\,v_{Ain}$ against $k\,x_0$ for various values of normalized Hall scale $H$. We see that when the Hall scale is comparable to the typical size of the system, i.e. $ H = 1$, only long wavelength sausage wave can propagate in the medium. Since Hall scale is dependent on the fractional ionization $\sim 1 / \sqrt{X_e}$ (Eq.~(\ref{eq:HSL})), this implies that unlike highly ionized medium, in a weakly ionized medium the cut—-off of the sausage surface waves depend on the fractional ionization. These cut—-offs are determined by Eq.~(\ref{eq:scd}). The decreasing cut—-off with increasing $k\,x_0$ implies decreasing Hall scale since $k x_0 \sim 1 / H$.  The decrease in the Hall scale in turn implies increasing ionization. As a result increasingly shorter wavelength sausage and kink wave will propagate with increasing altitude in the solar atmosphere. 

Whereas in the purely Hall case, the phase velocity of the sausage wave can become arbitrarily large in the vicinity of the cut—-offs, it is seldom the case in the partially ionized lower solar atmosphere where no matter how small, the presence of Pedersen diffusion \citep{P08b, PW12, PW13} will cause the damping of the waves. 

\section{Discussion and summary}
The solar atmosphere is partially ionized with Ohm, Hall and ambipolar diffusion playing important role at different scale heights \citep{PW13}. The flux tubes are formed at the granular boundaries through the interaction of the gravity with the stratified magnetic environment. The investigation of the surface waves in the partially ionized prominences suggest that the ambipolar diffusion will affect the short wavelength fluctuations whereas Ohmic diffusion will cause the damping of the long wavelength fluctuations \citep{S09}. Present investigation of the waves in the planer tube shows that both the ambipolar and Ohm diffusion causes the damping of the long wavelength fluctuations. In the purely Hall regime,  only long wavelength surface waves can propagate in the medium. This is related to the fact that the Hall effect introduces a scale in an otherwise scale free {\it ideal} medium and the propagation of the wave is inherently linked to this scale. Since the Hall scale is related to the fractional ionization, only sausage and kink surface waves whose wavelength is greater than the predetermined (by the level of ionization) threshold can propagate. In the photosphere-chromosphere, since all three diffusion are present \citep{PW13}, the surface wave will suffer damping. 

It is instructive to calculate the damping rate of the surface waves due to Pedersen diffusion. Assuming a weakly ionized solar plasma, we take $\eta_P / \eta_H \sim 10$, and $\rho_n \sim 10^{-8}\,\mbox{cm}^{-3}$ for the lower photosphere \citep{PW13}. Thus the damping rate, $\omega_i$ [Eq.~(\ref{eq:drt})] for $\rhox / \rhon = 0.25$ is of the order of  $\sim 400\,k\,v_{Ain}$. Assuming $B = 10^3\,\mbox{G}$, we get $ v_{A}\sim 10\,\mbox{cm}/\mbox{s}$. Thus the wave damping time is $t = 2\,\pi / \omega_i = \lambda / \left(4\times 10^3\right)$ where $\lambda = 2\,\pi / k$ is the wavelength of fluctuations. Therefore, for $\lambda \sim 10\,\mbox{km}$ ($1/10$ of the typical diameter of a thin tube), we see that the damping time is about four minutes. 

The Hall is the dominant diffusion mechanism in the entire photosphere--chromosphere in the weak field ($\sim 100\,\mbox{G}$) region (Fig.~1, Pandey $\&$ Wardle, 2013). The propagation of the sausage wave depends on the local value of the Hall scale length which in turn is determined by the fractional ionization of the plasma background. For example, when the Hall scale is comparable to the tube diameter, i.e. $ H = 1$, only long wavelength sausage wave can propagate in the medium. Since the decrease in Hall scale is due to increasing ionization, much shorter wavelength sausage and kink wave will propagate in the medium with increasing altitude in the solar atmosphere.

We shall note that the present model of piecewise constant density profile do not capture the resonant behaviour of the \alf surface waves, a process so vital to the heating of the coronal plasma \citep{G13}. Therefore the present model is illustrative in nature and will need to be generalized before the problem of coronal wave heating can be addressed. Namely, the continuum variation of the plasma parameters such as density or, magnetic field excites not the {\it line} but the continuum MHD spectrum which resonantly exchanges energy with the surrounding medium. Owing to the complexity of the problem, we have analysed only very simple case of the wave propagation in an incompressible medium and that too in various limiting cases. The plasma beta, which is a ratio of the plasma to the field pressure, varies in the solar photosphere \citep{GA01} and thus, the incompressibility assumption needs to be relaxed. 

Following is the summary of the present work.

1. The surface waves in the partially ionized medium is affected by the presence of the magnetic diffusion, although role of various diffusivities varies. 

2.  The phase velocity of the surface wave is not significantly modified in the presence of Hall diffusion, except only waves below certain cut—-off frequency can propagate in the medium. This cut-—off is tied to the Hall scale which in turn is dependent upon the fractional ionization of the medium. Thus, the fractional ionization of the ambient medium predetermines the nature of surface wave propagation. 

3. In the presence of both Hall and ambipolar diffusion, long wavelength fluctuations are damped in the medium.

4. The damping of the wave depends on the ambient plasma properties and also on the strength of the magnetic field.

\section*{Acknowledgments}
{It is my great pleasure to thank Mark Wardle for many discussions on this   subject. The financial support of the Australian Research Council through grant DP 130104873 is gratefully acknowledged. This research has made use of NASA's Astrophysics Data System.}

\end{document}